\documentclass[acmsmall,authorversion]{acmart}

\usepackage{xcolor}
\usepackage{algorithm}
\usepackage{algpseudocode}
\usepackage{multirow}
\usepackage{appendix}
\usepackage{tabularx}
\usepackage{booktabs} 
\usepackage{enumitem} 
\usepackage{pdflscape}
\usepackage{pgfplots}
\usepgfplotslibrary{statistics}
\usepackage{pgfplots}
\pgfplotsset{compat=1.17}

\algdef{SE}[INDENT]{Indent}{EndIndent}{\State \algorithmicindent}{} 
\algtext*{Indent}
\algtext*{EndIndent}

\AtBeginDocument{%
  }

\received{November 2023}
\received[revised]{January 2024}
\received[accepted]{March 2024}

\begin{document}

\setcopyright{acmlicensed}
\acmJournal{PACMHCI}
\acmYear{2024} \acmVolume{8} \acmNumber{ETRA} \acmArticle{231} \acmMonth{5}\acmDOI{10.1145/3655605}

\title{Privacy-preserving Scanpath Comparison for Pervasive Eye Tracking}

\author{Suleyman Ozdel}
\affiliation{%
  \institution{Technical University of Munich}
  \city{Munich}
  \country{Germany}
}
\email{ozdelsuleyman@tum.de}
\orcid{0000-0002-3390-6154}

\author{Efe Bozkir}
\affiliation{%
  \institution{Technical University of Munich}
  \city{Munich}
  \country{Germany}}
\email{efe.bozkir@tum.de}
\affiliation{%
  \institution{University of Tübingen}
  \city{Tübingen}
  \country{Germany}}
\email{efe.bozkir@uni-tuebingen.de}
\orcid{0000-0002-4594-4318}

\author{Enkelejda Kasneci}
\affiliation{%
  \institution{Technical University of Munich}
  \city{Munich}
  \country{Germany}}
\email{enkelejda.kasneci@tum.de}
\orcid{0000-0003-3146-4484}

\renewcommand{\shortauthors}{Ozdel et al.}

\begin{abstract}
As eye tracking becomes pervasive with screen-based devices and head-mounted displays, privacy concerns regarding eye-tracking data have escalated. While state-of-the-art approaches for privacy-preserving eye tracking mostly involve differential privacy and empirical data manipulations, previous research has not focused on methods for scanpaths. We introduce a novel privacy-preserving scanpath comparison protocol designed for the widely used Needleman-Wunsch algorithm, a generalized version of the edit distance algorithm. Particularly, by incorporating the Paillier homomorphic encryption scheme, our protocol ensures that no private information is revealed. Furthermore, we introduce a random processing strategy and a multi-layered masking method to obfuscate the values while preserving the original order of encrypted editing operation costs. This minimizes communication overhead, requiring a single communication round for each iteration of the Needleman-Wunsch process. We demonstrate the efficiency and applicability of our protocol on three publicly available datasets with comprehensive computational performance analyses and make our source code publicly accessible.
\end{abstract}




\maketitle

\section{Introduction}
In the rapidly evolving landscape of interactive technologies, eye tracking has been integrated into various devices, ranging from traditional stationary equipment to virtual reality (VR) headsets and smart glasses. This integration strives to refine intelligent user interfaces and yields insights into user visual behavior. Scanpaths, the sequential representations of eye movements, are utilized for gaze pattern analyses, providing a wealth of information about personal characteristics, such as skills expertise~\cite{castner_etal_2018}, health status~\cite{eraslan_etal_2020,harezlak2018application,avital2015method}, decision making behaviors~\cite{Zhou_Etal_2016}, sexual preferences~\cite{LAENG2007520}, and race~\cite{bar_haim_etal_2006}, to count a few. Such personal characteristics often include sensitive data, leading to the need for privacy considerations when handling scanpath data~\cite{liebling_preibusch_2014, bozkir2023eyetracked}. 

To encode scanpath data, different representations were employed~\cite{fahimi2021metrics,anderson2015comparison}, ranging from the direct numerical coordinates to saliency maps~\cite{borji2012state} and string sequences~\cite{noton1971scanpaths}. Among these, string-based representations are widely used and to assess similarities in string-encoded scanpaths, alignment techniques, particularly the edit distance (i.e., Levenshtein distance), are utilized~\cite{brandt1997spontaneous,anderson2015comparison,west2006eyepatterns,josephson2002attention}. In this context, the Needleman-Wunsch algorithm~\cite{needleman1970general} is recognized for its extensive use in comparative analyses~\cite{brandt1997spontaneous,cristino2010scanmatch,anderson2015comparison}. 

Given the extensive use of sensitive eye-tracking data across various fields, as previously outlined, developing robust, privacy-preserving string alignment algorithms designed for scanpath comparison is essential. Secure computation methods for string sequence alignment, prevalent in genomics~\cite{rane2010privacy,wang2015efficient,aziz2017secure}, are not commonly applied to eye-tracking data. These methods, often involving a third-party intermediary, are primarily optimized for DNA query search rather than obtaining an exact similarity score~\cite{kantarcioglu2008cryptographic,wang2015efficient,aziz2017secure,asharov2017privacy}. Furthermore, although there are secure protocols designed to obtain exact similarity scores, some rely on computationally intensive fully homomorphic encryption schemes~\cite{cheon2015homomorphic}. In contrast, others necessitate significant communication overhead between involved parties and use various protocols depending on the substitution cost definitions~\cite{rane2010privacy, atallah2003secure}.

The existing literature indicates a significant gap in developing computationally efficient and practical secure two-party string alignment protocols, specifically in the context of scanpaths. Creating these protocols is crucial for the secure and private analysis of eye-tracking data, a need that is becoming more pronounced with the increasing use of eye-tracking devices. We introduce a novel two-party secure string alignment protocol to bridge this gap, specifically for scanpath comparisons. This protocol is intricately designed for the Needleman-Wunsch algorithm and also offers the flexibility to be utilized for other edit distance algorithms, thereby expanding its applicability to DNA sequence analysis. Our protocol supports various substitution cost definitions and minimizes inter-party communication. Furthermore, it utilizes the Paillier additive homomorphic encryption scheme, chosen for its relatively lower computational demands compared to fully homomorphic encryption schemes, to enable secure computations. In summary, our work introduces a novel approach to enhance privacy and efficiency in scanpath comparisons, with the following five main contributions:
\begin{itemize}
    \item We introduce the first-ever method dedicated to securing privacy in the comparison of scanpaths, signifying a pioneering advancement in eye tracking. 
    \item We propose an efficient two-party computation (2PC) protocol for scanpath comparison requiring only a single round of communication between parties and is applicable to the edit distance kind string alignment algorithms.
    \item We introduce a novel probabilistic matrix processing strategy for the Needleman-Wunsch algorithm that conceals the computation of specific cells from another party to enhance security.
    \item We introduce a masking technique incorporating order-preserving masking by exploiting the Paillier cryptosystem's properties to ensure the privacy of minimum cost computation in the Needleman-Wunsch algorithm.
    \item We show the practical applicability and effectiveness of our method by evaluating it on three publicly available eye-tracking datasets and make our source code publicly accessible for reproducibility and transparency.
\end{itemize}

\section{Related Work}
We discuss the previous research in two lines of work, namely, privacy-preserving eye tracking in Section~\ref{label_subsec_ppet} and privacy-preserving string comparison in Section~\ref{lbl_subsec_ppsaa}, as our work focuses on strings for scanpath comparison. 

\subsection{Privacy-preserving Eye Tracking}
\label{label_subsec_ppet}
Eye gaze and pupillometry provide beneficial information in various applications, especially for visual interaction, as incorporating eye-tracking data can facilitate hands-free interaction. However, it is known that the same data combined with the presented visual stimulus can reveal sensitive information about humans~\cite{liebling_preibusch_2014, Kroeger_etal_2020}. To count a few, previous work found that eye-tracking data is related to sexual preference~\cite{wenzlaff_etal_2016}, body mass index~\cite{GRAHAM2011577}, health status~\cite{WILLIAMS2010617}, and personal identifiers~\cite{bozkir_diff_privaccy2021} when relevant stimulus is encountered. Considering these, the importance of privacy protection for eye-tracking data has constantly been emphasized in the context of visual analytics~\cite{silva_etal_2019}, security applications~\cite{katsini_etal_2020}, virtual reality~\cite{bozkir2023eyetracked}, and pervasive computing~\cite{gressel_2023_why}. Yet few works indeed proposed technical approaches to protect privacy. 

Differential privacy, a privacy protection method that focuses on the privacy risk of an individual participating in a database, has recently been utilized on different forms of eye-tracking data. For instance,~\citeauthor{liu_etal_2019} ~\cite{liu_etal_2019} utilized the Gaussian mechanism of differential privacy on heatmaps whereas~\citeauthor{steil_etal_2021}~\cite{steil_etal_2021} applied its exponential mechanism to aggregated eye movement features to protect privacy. However, standard differential privacy mechanisms are vulnerable to the correlations in the data. To address this issue,~\citeauthor{bozkir_diff_privaccy2021}~\cite{bozkir_diff_privaccy2021} took temporal correlations in eye movements into account and utilized differential privacy by decorrelating the data in the frequency domain.With a similar aim,~\citeauthor{kaleido_2021}~\cite{kaleido_2021} provided privacy protection to eye-tracking data by considering spatio-temporal attacks on gaze data streams with a method that utilizes differential privacy. However, differential privacy achieves privacy protection by adding a significant amount of randomly generated noise, and such noise often leads to a certain amount of performance reduction in utility tasks; therefore, achieving an optimal privacy-utility trade-off is usually challenging. In addition, standard mechanisms of differential privacy are vulnerable to correlations in the data, and as eye-tracking data is highly correlated, particularly in the temporal direction, which is another challenge to address when differential privacy mechanisms are utilized for privacy protection. Previous work also focused on other notions of privacy for eye-tracking data, such as k-anonymity and plausible deniability together with differential privacy~\cite{brendan_kanonymity_2022, brendan_holistic_tvcg_2023} and found that while plausible deniability and differential privacy provide practical privacy-utility trade-offs, k-anonymity performs the best at gaze prediction utility task. 

Due to the aforementioned challenges, other research focused on more practical approaches to address privacy issues in pervasive eye tracking. For instance, ~\citeauthor{subsampling_brendan_tvcg_2021}~\cite{subsampling_brendan_tvcg_2021} proposed spatial and temporal downsampling in the eye-tracking data and showed that person re-identification rates drop significantly when their method is applied, while utility tasks work with reasonable performance. Similarly, ~\citeauthor{fuhl_2021}~\cite{fuhl_2021} utilized a reinforcement learning-based approach by treating subject and gender information as protected while document-type and expertise classification tasks as utility tasks. The authors showed that their approach outperforms differential privacy- and generative adversarial network-based solutions for protecting privacy yet providing privacy protection probabilistically. ~\citeauthor{pmlr_v210_elfares23a}~\cite{pmlr_v210_elfares23a} focused on federated learning for gaze estimation in the wild and showed that their approach outperforms vanilla federated learning in this task. Yet, most of these works either add a significant amount of noise in the data or work probabilistically, which is questionable from a regulation point of view. To this end, ~\citeauthor{ppge_bozkir_unal_2020}~\cite{ppge_bozkir_unal_2020} utilized a randomized encoding-based framework to provide formal privacy guarantees for the gaze estimation task, where two-input parties provide their data to train a gaze estimation model on a cloud without revealing their sensitive eye movement data. As it is possible to utilize such formal guarantees potentially in an efficient way, as indicated by previous work, we also argue for formal methods to protect privacy in the scanpath comparison task. 

\subsection{Privacy-preserving String Alignment Algorithms}
\label{lbl_subsec_ppsaa}
String alignment algorithms are essential for analyzing similarities in both scanpaths and DNA sequences. Due to the sensitive nature of this information, the development of privacy-preserving string alignment algorithms is essential in this context. There are several works~\cite{aziz2017secure,asharov2017privacy,schneider2019episode,zheng2019efficient,kantarcioglu2008cryptographic} that mainly focus on DNA query search, often utilizing private protocols for edit distance approximations. In contrast to the aforementioned works focusing on query search, ~\citeauthor{jha2008towards}~\cite{jha2008towards} utilized Yao's garbled circuits~\cite{yao1986generate} for private edit distance computation between two parties and additionally proposed an alternative protocol to compute Smith-Waterman score \cite{smith1981identification}. ~\citeauthor{zhu2020efficient}~\cite{zhu2020efficient} also employed Garbled Circuits for private computation of edit distance, addressing both semi-honest and malicious adversary models. ~\citeauthor{ayday2013protecting}~\cite{ayday2013protecting} introduces a framework utilizing a modified Paillier cryptosystem for the secure storage and processing of patient genomic data, allowing medical centers to process genome data privately.

Moreover, several methods~\cite{atallah2003secure,rane2010privacy,cheon2015homomorphic} leverage homomorphic encryption to enhance privacy. ~\citeauthor{atallah2003secure}~\cite{atallah2003secure} introduced a technique to determine sequence similarity through a two-party secure computation protocol. Their approach harnesses homomorphic encryption, storing the alignment matrix under additive sharing between the two parties. A key aspect of this approach is the minimum-finding protocol, which requires two communication rounds between parties and is used three times per cell computation, increasing the communication overhead. Similarly, ~\citeauthor{rane2010privacy}~\cite{rane2010privacy} proposed an asymmetric two-party computation protocol tailored for server-client interactions, which also leverages additive secret sharing and homomorphic encryption. ~\citeauthor{madrigal2023privacy}~\cite{madrigal2023privacy} proposed an algorithm based on secret sharing for DNA comparison, leveraging the Wagner-Fischer edit distance to achieve reduced execution times in comparison tasks under both passive and active security scenarios. ~\citeauthor{yoshimoto2020faster}~\cite{yoshimoto2020faster} proposed a homomorphic encryption-based two-party secure computation protocol for the modified edit distance with moves algorithm aimed at reducing the round complexity. ~\citeauthor{cheon2015homomorphic}~\cite{cheon2015homomorphic} presented an approach for the private computation of edit distance employing somewhat homomorphic encryption. Their framework, which employs specific circuits for equality, comparison, and addition, leverages a third party, typically a cloud server, to perform computations on encrypted data, thereby ensuring data confidentiality.  

Existing string alignment methods are mainly developed for genome analysis, typically dealing with a small alphabet of four letters. Most of them~\cite{aziz2017secure,asharov2017privacy,schneider2019episode,zheng2019efficient,kantarcioglu2008cryptographic} are developed for query search, using approximations and often include a third party in their protocols. Methods applicable for two-party computation are not computationally efficient~\cite{cheon2015homomorphic} and usually necessitate frequent communication between parties and utilize different protocols tailored to different substitution costs~\cite{rane2010privacy,atallah2003secure}. Thus, a research gap remains in achieving an efficient two-party secure string alignment computation that balances computation time and communication load. To address this gap, our protocol is designed to accommodate a range of substitution costs and to calculate the Needleman-Wunsch algorithm, which is a generalized version of edit distance. Its primary benefit lies in its efficiency, requiring only a single round of communication between parties per iteration, thereby significantly simplifying the process, and it supports scanpath comparisons based on eye movement data, which is highly missing in the eye-tracking literature.

\section{Methods}

To provide privacy-preserving scanpath comparisons, we present a novel secure two-party computation protocol to privately compute the Needleman-Wunsch algorithm between two parties without the involvement of a third-party entity, such as a cloud instance. Utilizing the Paillier homomorphic encryption scheme, which offers the necessary properties such as ciphertext addition, scalar multiplication, and probabilistic encryption, our approach executes the Needleman-Wunsch algorithm within the encrypted domain. Our method ensures that no information about the individual scanpaths is disclosed except for their lengths and the final similarity value. 

\subsection{Preliminaries}
Before diving into the specifics of our methodology, we provide a concise overview of the foundational concepts that anchor our protocol. Two critical components form the backbone of our method: the Needleman-Wunsch algorithm and the Paillier homomorphic encryption scheme. The Needleman-Wunsch algorithm is one of the fundamental methods to compare and derive similarity scores for two scanpaths. On the other hand, the Paillier homomorphic encryption scheme offers unique cryptographic properties that permit mathematical operations in the encrypted domain. This section introduces these core concepts to furnish the reader with the requisite background knowledge. 

\subsubsection{Needleman-Wunsch Algorithm}
The Needleman-Wunsch algorithm~\cite{needleman1970general} stands as a generalized version of the Edit distance algorithm, also known as Levenshtein distance~\cite{levenshtein1966binary}. Its primary objective is to determine the best global alignment between two sequences, achieving either maximum similarity or minimum dissimilarity. To achieve this, the algorithm uses a scoring system that considers matches, mismatches, insertions, and deletions as part of its calculation.

The Needleman-Wunsch algorithm was first introduced for comparing DNA or protein sequences~\cite{needleman1970general}, and it found its application in eye tracking to align scanpaths, enabling the comparative analysis of eye movement patterns~\cite{castner2018scanpath,cristino2010scanmatch}. The final alignment score generated by the Needleman-Wunsch algorithm serves as a crucial metric for assessing the similarity of sequences or gaze data. This alignment score helps identify shared or distinct aspects of visual attention among individuals or in response to different stimuli~\cite{castner2018scanpath, cristino2010scanmatch, day2010examining, busjahn2015eye}.


Let \( M \) be defined as an alignment matrix with dimensions \( m \times n \), which represents the cost associated with aligning two sequences up to the \( i^{th} \) and \( j^{th} \) positions, respectively. The matrix \( M \) is initialized as follows:
\[
\begin{array}{lll}
M(0,0) = 0, & M(i,0) = i \times c_{del} \quad \text{for } 1 \leq i \leq m, & M(0,j) = j \times c_{ins} \quad \text{for } 1 \leq j \leq n, 
\end{array}
\]
where $c_{del}$ and $c_{ins}$ represent the costs of deletion and insertion, respectively. For other values where \( 1 \leq i \leq m \) and \( 1 \leq j \leq n \), \( M \) is defined as:
\[
M(i,j) = \min \left\{ 
\begin{array}{lcl}
M(i-1,j-1) + S(\lambda_i, \mu_j), & M(i-1,j) + c_{del}, & M(i,j-1) + c_{ins}
\end{array}
\right\},
\]
with \( S(\lambda_i, \mu_j) \) denoting the substitution cost between the letters \( \lambda_i \) and \( \mu_j \).

In the alignment matrix \( M \), for the computation of \( M(i,j) \), it is needed to have the three previous entries: \( M(i-1,j-1) \), \( M(i-1,j) \), and \( M(i,j-1) \). As long as these entries are available, there is no strict requirement to follow a specific order, even though many dynamic programming algorithms traditionally proceed in a row-by-row or column-by-column fashion. Once the dynamic programming is done and all cells in the matrix are filled, the entry \( M(m,n) \) gives us the final alignment used as a similarity metric. The overall time complexity of this algorithm is therefore \( \mathcal{O}(mn) \). The pseudocode of this algorithm is given in the Appendix \ref{app:pseudocode-NeedlemanWunch}.

\subsubsection{Paillier Cryptosystem}
Paillier encryption scheme is a semantically secure asymmetric homomorphic encryption scheme that enables data sharing and processing without revealing the underlying content. Semantically secure algorithms maintain their security even when an adversary can access pairs of messages (i.e., plaintexts) and their associated encrypted messages (i.e., ciphertexts). Asymmetric encryption systems work with a dual-key setup: the public key facilitates encryption and homomorphic operations, while the private key is essential for decryption. This arrangement allows a third party to compute operations on the encrypted data using only the public key. The key generation protocol in the Paillier cryptosystem accepts a security parameter that specifies the number of bits for the prime numbers used in the key creation. A larger bit length results in more secure keys by increasing the difficulty of potential cryptographic attacks, but it also raises the computational requirements.

The Paillier encryption scheme uses probabilistic encryption, which ensures that a given plaintext maps to many possible ciphertexts, providing a high level of security by making it computationally infeasible for an attacker to deduce the plaintext from the ciphertext, even when the same plaintext is encrypted multiple times. Additionally, it exhibits homomorphic properties that allow certain operations on ciphertexts, such as the addition of ciphertexts, which corresponds to the addition of their plaintexts, and the multiplication of ciphertext by an unencrypted scalar, equating to the multiplication of the plaintext by that scalar. Subtraction can be performed using the additive inverse in the encrypted domain. While direct division is not supported, division by a plaintext scalar can be achieved by multiplying with its multiplicative inverse. Further details are given in Appendix~\ref{Appendix:Paillier Encryption Scheme}.

For our proposed protocol, the encryption mechanism must accommodate distinct operations, namely, adding ciphertexts and multiplication either with a scalar value (i.e., an unencrypted number) or with ciphertexts. Additionally, probabilistic encryption, which yields varied ciphertexts for a single plaintext, is a key feature for privately computing the Needleman-Wunsch algorithm. Although fully homomorphic encryption (FHE) schemes also support these operations, they require significant computational demands and larger ciphertexts for the same security level, leading to increased bandwidth consumption. Consequently, considering these, we selected the Paillier cryptographic system~\cite{paillier1999public}, which inherently has the required capabilities.

\subsection{Framework}
In this section, we discuss our framework for privacy-preserving scanpath comparison. In our framework, there are two primary actors, namely Alice and Bob, who might be individuals or patients looking to compare their scanpaths. Alice takes on the key holder role, possessing both the secret and public key pairs generated using the Paillier cryptosystem. In contrast, Bob acquires the public key and uses it to run the Needleman-Wunsch algorithm on the encrypted domain.

\paragraph{Threat Model.} In our proposed model, we engage with two parties aiming to compare scanpaths. This interaction operates under the assumption of a ``semi-honest'' behavior from both entities. The semi-honest model, often called the ``honest-but-curious'' model, describes participants in cryptographic schemes who strictly follow the given protocol. They do not deviate from the provided steps or change the process. However, they are naturally curious. While they stick to the rules, they try to learn any extra information from the eye-tracking data they observe during the protocol's operation. In simple terms, these participants act as instructed but are always keen to gather sensitive information from others' eye-tracking data without actively interfering.

\paragraph{Masking Process for Minimum Cost Computation.}\label{par: Masking Process for Minimum Cost Computation}
Before the descriptions of our scanpath comparison protocol, we first introduce the masking process required in each iteration. To execute the Needleman-Wunsch algorithm, finding the minimum among the sequence of editing operation costs (insertions, deletions, or substitutions) in each iteration is essential. However, Bob performs computations on encrypted data, and all these costs are encrypted. As ciphertexts do not reveal any information about their corresponding plaintexts, it is impossible to determine the minimum of these encrypted values without involving the owner of the secret key. Therefore, we must interact with Alice, the owner of the private key, to compute the minimum by decrypting these values. Alice could discern each value if she knew the current step and the vector. To address this issue, the given vector is masked and subsequently permuted to obscure the information from Alice. Initially, an order-preserving masking is applied, followed by an affine transformation, and finally, the values are permuted.

We propose an order-preserving masking approach that preserves the original sequence's order, enabling the retrieval of initial values through a uniform method applicable across various variables. Initially, we have the vector \(\mathbf{x} = [x_1, x_2, \ldots, x_m]\). This mechanism operates as follows. For each element \(x_i\) in the vector \(\mathbf{x}\), the updated value \(x_i'\) is computed using the formula:
\begin{equation}
x_i' = (x_i \cdot \rho_1) - \sum_{\substack{j=1 \\ j \neq i}}^{m} x_j, 
\end{equation}\label{eq:order_preserving_masking}
where \( \rho_1 + 1 \) has a multiplicative inverse in modular \( n \) and \( \rho_1 > 0 \). The proof is given in Appendix~\ref{Appendix:Proof Order Preserving Masking}.

To obscure the data from Alice, Bob applies this masking with a probability specified in \text{\ref{step: 2}}. However, the retrieval of the original value is also essential. The inverse function is achieved by subtracting the sum of all initial values in the vector from the masked value and then multiplying the resulting sum with the multiplicative inverse of \( (\rho_1 + 1) \). This is mathematically represented as:
\[
x_i = \left( x_i' + \sum_{j = 1}^{m} x_j \right) \times (\rho_1 + 1)^{-1}.
\]

Subsequently, an affine transformation is applied, utilizing three random variables—two for addition and one for multiplication—to mask the actual values given in \ref{step: 3}. After receiving the minimum value from Alice, Bob needs to retrieve the original value by applying the inverse transformation using subtraction and the multiplicative inverse. Following this, a random permutation is employed to obfuscate the sequence of values, ensuring that Alice cannot determine whether the minimum value resulted from substitution, deletion, or insertion costs. Bob only receives the minimum value and does not require reversing this permutation. 

\paragraph{Privacy-preserving Needleman-Wunsch protocol.}  After the aforementioned operations, in the following, we provide a comprehensive protocol overview
by first presenting the essential definitions and notations that underpin our framework in Table \ref{tab: definitions}. A detailed visual flow of our protocol is depicted in Figure~\ref{fig:sequence_diagram} in the Appendix~\ref{app:sequence_diagram}. Additionally, the pseudocode for each part of the algorithm is provided in Appendix \ref{app:Pseudocode Privacy Preserving Protocol}.

\begin{table}[h]
\centering
\caption{Definitions of symbols used in the algorithm.}
{\footnotesize 
\begin{minipage}[t]{0.49\textwidth}
\begin{tabularx}{\textwidth}{cX}
\toprule
\textbf{Symbol} & \textbf{Definition} \\
\midrule
\( \kappa \) & Security parameter for the Paillier cryptosystem, representing the bit length of the keys. \\
\( \mathcal{E}_{pk}(p) \) & Encryption of plaintext \( p \) with public key \( pk \). \\
\( \mathcal{D}_{sk}(c) \) & Decryption of ciphertext \( c \) with secret key \( sk \). \\
\( \mathbf{s_A} \) & Alice's scanpath vector of size \( m \). \\
\( \mathbf{s_B} \) & Bob's scanpath vector of size \( n \). \\
\( \mathbf{\alpha} \) & The alphabet vector, e.g., \( \mathbf{\alpha} = [A, B, \ldots, Z, a, b, \ldots, z] \). \\
\( \mathbf{D} \) & A matrix of size \( n \times |\mathbf{\alpha}| \), where \( D_{i,j} \) denotes the encrypted distance value between the \( i \)-th element of \( \mathbf{s_A} \) and the \( j \)-th letter in \( \mathbf{\alpha} \). \\
\bottomrule
\end{tabularx}
\end{minipage}
\hfill 
\begin{minipage}[t]{0.49\textwidth}
\raisebox{0pt}{ 
\begin{tabularx}{\textwidth}{cX}
\toprule
\textbf{Symbol} & \textbf{Definition} \\
\midrule
\( \mathbf{k_B} \) & Vector such that each element \( k_{B_j} \) represents the index of \( s_B(j) \) in the alphabet \( \mathbf{\alpha} \), where \( k_{B_j} = \mathbf{\alpha}^{-1}(s_B(j)) \). \\
\( \mathbf{C} \) & Candidate vector. Contains indices \( (i, j) \) for elements in \( M \) pending computation where dependent values \( M(i-1,j-1) \), \( M(i-1,j) \), and \( M(i,j-1) \) are already computed. \\
\( c_{ins}\) & Insertion cost. \\
\( c_{del}\) & Deletion cost. \\
\( \otimes \) & Scalar multiplication with a ciphertext. \\
\( \oplus \) & Addition operation for two ciphertexts. \\
\addlinespace[4pt] 
\bottomrule
\end{tabularx}
}
\end{minipage}
}
\label{tab: definitions}
\end{table}

Firstly, to compute the distance privately, we must set up the cryptographic framework, ensure secure data sharing, and initialize the requisite variables for the primary protocol.

\textit{Setup:}
\begin{enumerate}[label=\roman*.]
\item Alice generates a secret and public key pair \( (sk, pk) \) using the Paillier cryptosystem with security parameter \(\kappa\). Subsequently, Alice shares \( pk \) with Bob for further cryptographic computations.
 \end{enumerate}

\textit{Initialization:}
\begin{enumerate}[label=\roman*.]
\item Alice constructs the substitution cost matrix \( \mathbf{D} \), where each element \( D_{i,j} \) represents the encrypted substitution cost between the \( i^{th} \) element of her scanpath vector \( \mathbf{s_A} \) and the \( j^{th} \) letter in the alphabet \( \mathbf{\alpha} \). Specifically, \( D_{i,j} = \mathcal{E}_{pk}\big(S(s_{A_i}, \alpha_j)\big) \) for all \( 1 \leq i \leq n \) and \( 1 \leq j \leq |\mathbf{\alpha}| \), where \( S(s_{A_i}, \alpha_j) \) denoting the substitution cost between the letters \( s_{A_i} \) and \( \alpha_j \). Following this, Alice sends the encrypted distance matrix \( \mathbf{D} \) to Bob.

\item Bob initializes the alignment matrix \( M \) for the Needleman-Wunsch algorithm in encrypted form:
        \begin{align*}
            M_{0, 0} &= \mathcal{E}_{pk}(m_{0,0}), \\
            \forall i > 0, \, M_{i, 0} &= \mathcal{E}_{pk}(m_{i, 0}), \\
            \forall j > 0, \, M_{0, j} &= \mathcal{E}_{pk}(m_{0, j}),
        \end{align*}
        where \( m_{i, 0} = \sum_{k=1}^{i} c_{del}\) and \( m_{0, j} = \sum_{k=1}^{j} c_{ins}\).
 \end{enumerate}

Upon completing the setup and initialization phase, Bob starts computing the remaining values in the alignment matrix \( M \). To fill \( M \), each value for the pairs \( (i,j) \), where \(i \in [1, m]\) and \(j \in [1, n]\), needs to be computed. The computation is executed in a random order, instead of following the conventional dynamic programming order, to effectively obscure the current step from Alice.

\textbf{Candidate Vector Construction}: A candidate vector, denoted as \( \mathbf{C} \), contains the indices \( (i, j) \) for the elements in \( M \) that are pending computation and for which the dependent values \( M(i-1,j-1) \), \( M(i-1,j) \), and \( M(i,j-1) \) are already computed. In each iteration, this candidate vector is updated by checking the dependent values for possible candidates \( M(i+1,j+1) \), \( M(i+1,j) \), and \( M(i,j+1) \).

To ensure randomness in the computation, a pair of indices \( (i, j) \) is randomly selected from \( \mathbf{C} \) in each step, and the corresponding cell is computed. At the outset, given that the initial conditions of \( M \) have been established, \( \mathbf{C} \) contains only the index \( (1,1) \). 

We present a step-by-step description of the operations executed within each iteration loop.
\begin{enumerate}[label=\textit{Step \arabic*}]
\item  According to randomly selected indices  \( (i, j) \), Bob computes encrypted editing operation costs as follows:
    \begin{align*}
    x_1 &= M(i-1, j-1) \oplus D(i, \mathbf{k_B}[j-1]), \\
    x_2 &= M(i, j-1) \oplus \mathcal{E}_{pk}(c_{ins}), \\
    x_3 &= M(i-1, j) \oplus \mathcal{E}_{pk}(c_{del}).
    \end{align*}
After computing these values, Bob aggregates them into a vector, denoted as \( \mathbf{x} = [x_1, x_2, x_3] \).
\item \textit{Order Preserving Masking:}\label{step: 2} To introduce uncertainty and securely mask the data from Alice, Bob randomly selects one of the following two approaches, each with a probability of 0.5: 
    \begin{enumerate}[label=\textit{Option }\arabic*.]
        \item Multiply by a random number that has a multiplicative inverse and let \( x_{\ell}' = x_{\ell} \otimes \rho_1 \), where \(\ell \in \{1,2,3\}\).
        \item Apply an order-preserving mask introduced in the previous masking process. Bob randomly selects a value \( \rho_1 \) and ensures that \( \rho_1 + 1 \) has a multiplicative inverse within the defined domain. He then computes
        \begin{align*}
        x_1' &= (x_1 \otimes \rho_1) \oplus (-x_2) \oplus (-x_3), \\
        x_2' &= (x_2 \otimes \rho_1) \oplus (-x_1) \oplus (-x_3), \\
        x_3' &= (x_3) \otimes \rho_1) \oplus (-x_1) \oplus (-x_2). 
        \end{align*}
\end{enumerate}
\item \textit{Affine Transformation:} \label{step: 3} Bob masks \( \mathbf{x'}\) values using an affine transformation. He selects a value \( \rho_2 \), ensuring that \( \rho_2 \) has a multiplicative inverse in the given domain. Additionally, he randomly selects values \( \delta_1 \) and \( \delta_2 \). Subsequently, he applies the transformation as:
\begin{align*}
x_{\ell}'' &=  \Big(\rho_2 \otimes \big(x_{\ell}'  \oplus \mathcal{E}_{pk}(\delta_1)\big)\Big)  \oplus \mathcal{E}_{pk}(\delta_2), \quad \ell \in \{1,2,3\} \\
\mathbf{x''} &= [x_1'', x_2'', x_3''].
\end{align*}

\item \textit{Random Permutation:} Bob applies a random permutation order \(\pi\) to \(\mathbf{x''}\). He obtains the permuted vector \( \mathbf{x''_{\pi}} \) and transmits \( \mathbf{x''_{\pi}} \) to Alice.

\item \textit{Alice Minimum Cost Computation:} Alice decrypts the permuted values to determine the smallest value and then encrypts it as \(\mathcal{E}_{pk}(x^*)\), where \(x^* = \min\big(\mathcal{D}_{sk}(x''_{\pi(1)}), \mathcal{D}_{sk}(x''_{\pi(2)}), \mathcal{D}_{sk}(x''_{\pi(3)})\big)\). Alice will send an entirely different ciphertext due to the randomization in the Paillier cryptosystem encryption. Consequently, when Alice sends the encrypted value to Bob, Bob cannot discern which value corresponds to the minimum and which operation resulted in that minimum.

\item \textit{Bob Correction Operation:} Bob retrieves the value in the encrypted domain, and he first applies an inverse affine transform and obtains \(x_{\text{min}}'\) as:
\begin{equation}
x_{\text{min}}' = \left( \mathcal{E}_{pk}(m^*) \oplus \mathcal{E}_{pk}(-\delta_2) \right) \otimes \rho_2^{-1} \oplus \mathcal{E}_{pk}(-\delta_1).
\end{equation}
Then, if he did not apply the order-preserving mask, \(M(i, j) = x_{\text{min}}' \otimes \rho_1^{-1}\); otherwise, \(M(i, j)\) is calculated as:
\begin{equation}
M(i, j) = \big( x_{\text{min}}' \oplus (x_1 \oplus x_2 \oplus x_3 )^{-1}\big) \otimes (\rho_1 + 1)^{-1}.
\end{equation}
\end{enumerate}

All the Paillier encryption scheme operations are detailed in Appendix \ref{Appendix:Paillier Encryption Scheme}. In each computational iteration, a single value within the matrix is computed. The last computed value, \( \mathcal{E}_{pk}(M(m, n)) \), corresponds to the similarity score in the Needleman-Wunsch algorithm, indicating the similarity of scanpaths. To obtain the decrypted result, Bob must transmit this value to Alice. In turn, Alice employs her secret key to decrypt \( \mathcal{E}_{pk}(M(m, n)) \), yielding the ultimate result denoted as \( \Delta = \mathcal{D}_{sk}(\mathcal{E}_{pk}(M(m, n))) \). Subsequently, Alice conveys the decrypted result back to Bob.

\subsection{Security Analysis}

In our protocol, Bob receives two types of input from Alice: an encrypted distance matrix representing Alice's scanpath and an encrypted minimum element in each iteration. All inputs are encrypted using the Paillier scheme, ensuring that Bob cannot deduce any details about Alice's scanpath other than its length. Alice receives the vector for minimum cost computation in each step, which is the only kind of input she receives from Bob. To safeguard against potential information leaks during this transmission, we employed a probabilistic processing strategy and a masking method involving permutation. 

\begin{figure}[htbp]
    \centering
    \begin{minipage}{.48\textwidth}
        \centering
        \includegraphics[width=0.8\linewidth]{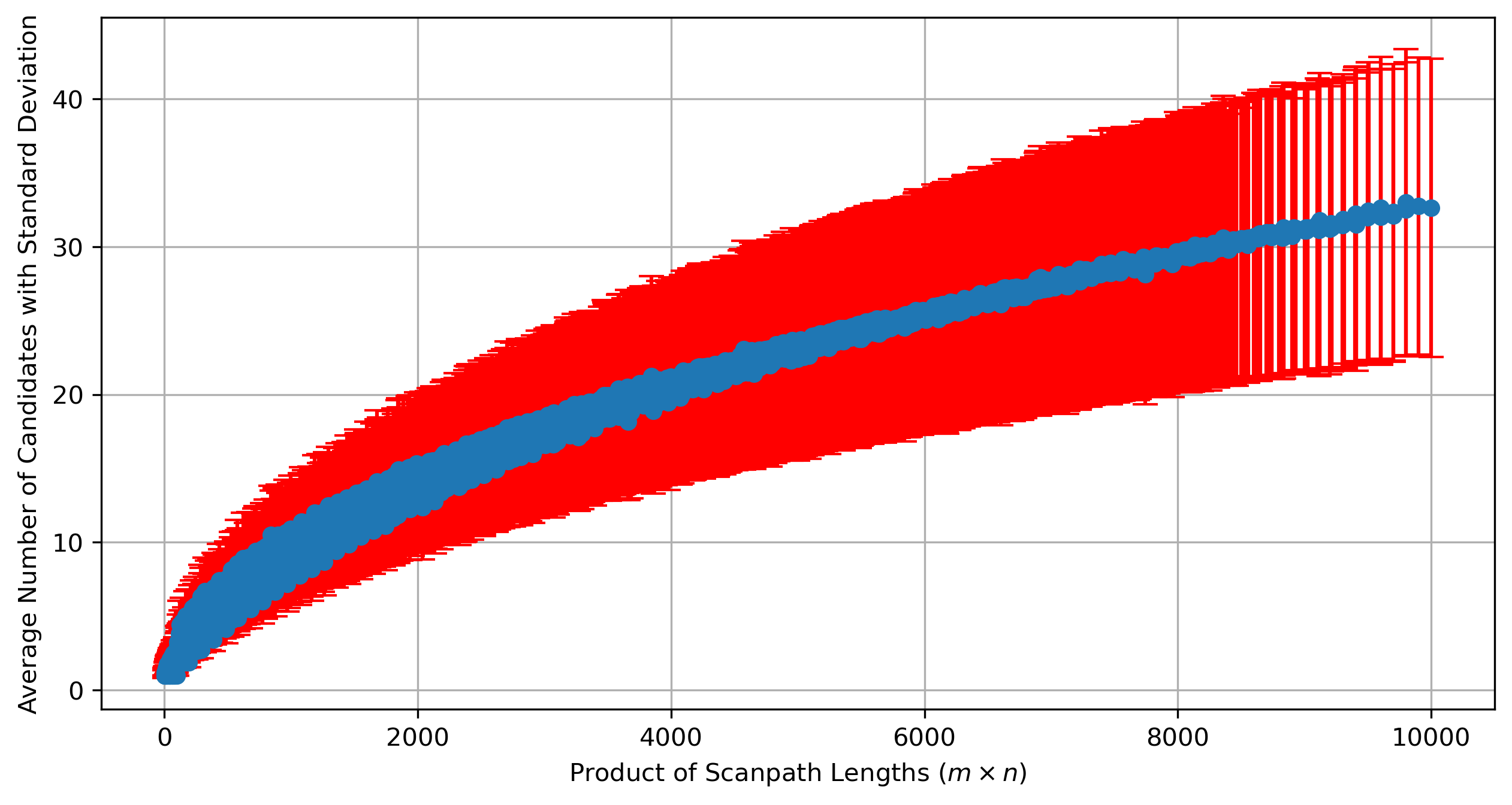} 
        \captionof{figure}{Relationship between the matrix size ($m \times n$) and the average number of candidate cells per iteration for the probabilistic Needleman-Wunsch algorithm.}
        \label{fig:mean_std}
    \end{minipage}\hfill
    \begin{minipage}{.48\textwidth}
        \centering
        \includegraphics[width=0.8\linewidth]{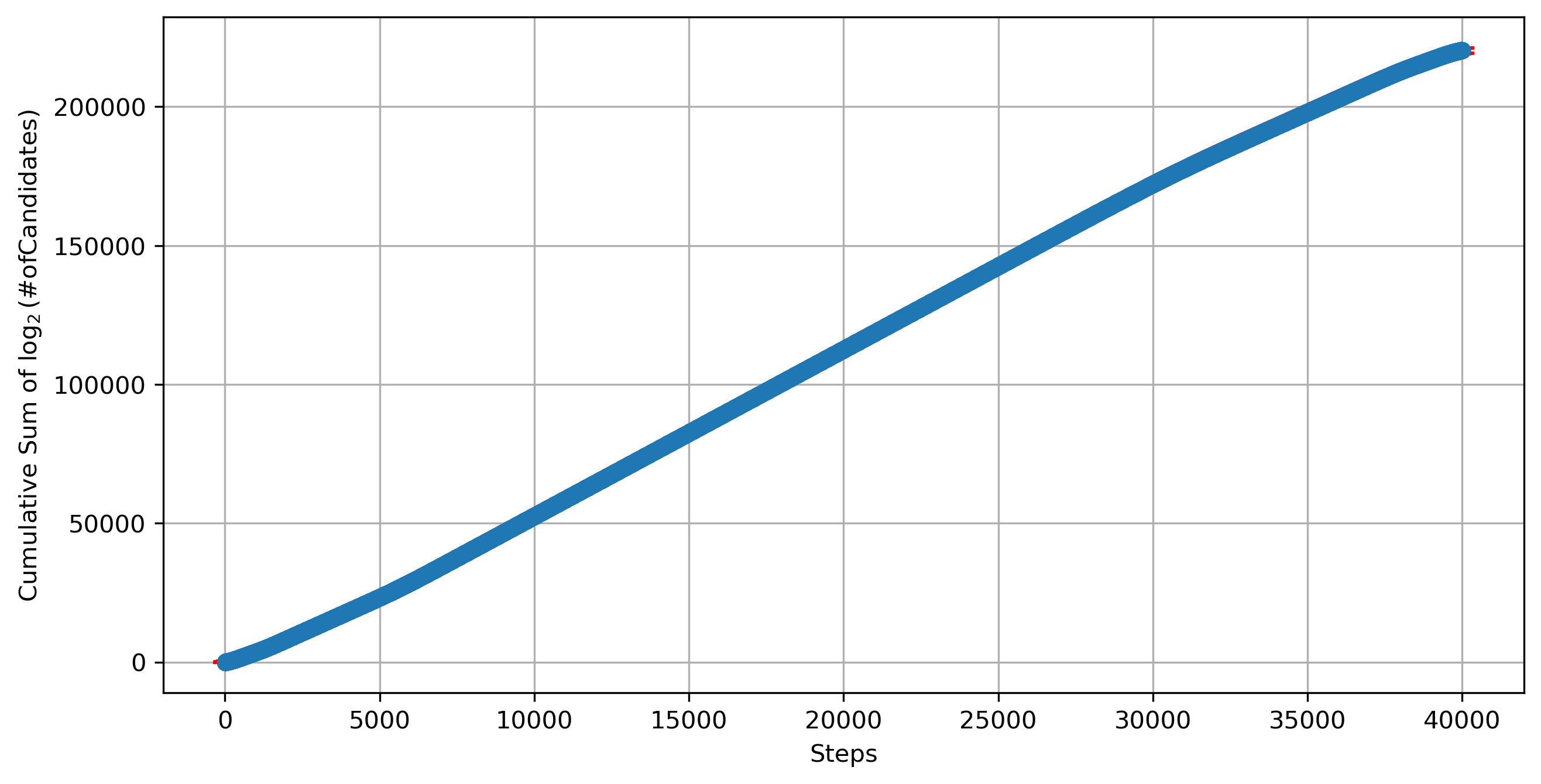} 
        \captionof{figure}{Cumulative computational complexity of the probabilistic selection process, represented by the sum of the $\log_2$ of the candidate counts per iteration for each step.}
        \Description{Cumulative computational complexity of the probabilistic selection process, represented by the sum of the $\log_2$ of the candidate counts per iteration for each step.}
        \label{fig:mean_product_base2}
    \end{minipage}
\end{figure}

In our protocol, Bob employs a probabilistic processing strategy to hide the current step from Alice. In each iteration, he randomly chooses a cell to process from the existing candidates, represented by \( \mathbf{C} \).  An example illustration of several steps of this process is provided in Figure~\ref{fig:matrix_processing} in the Appendix~\ref{app:matrix_processing}. The level of randomness in each iteration is associated with the number of candidates, which reflects the degree of uncertainty. This number of candidates depends on the current step and the length of scanpaths. Figure~\ref{fig:mean_std} demonstrates the relationship between the dimensions of the matrix ($m \times n$) and the average number of candidates, adding compounded complexity at each step. As the number of letters in the scanpaths increases, so does the average number of candidates, which enhances our security level. For illustrative purposes, consider scanpaths of lengths \(m = n = 20\). On average, $6.8$ candidates are observed in each iteration, as shown in Table~\ref{fig:mean_std}. This results in approximately \((6.8)^{400}\) or \(2^{1100}\) different ways to compute the matrix. Also, Figure~\ref{fig:mean_product_base2} illustrates the cumulative sum of the number of candidates in $\log_2$ for each iteration step while processing a $200 \times 200$ matrix. By step 46 ($i \times j$), it reaches 80, implying $2^{80}$ possible combinations, and by step $65$, it hits $2^{128}$ combinations. The $y$-axis represents the exponent in the base-2 logarithm. For instance, at step $10,000$, which may correspond to a matrix of $100 \times 100$ letters, there are $2^{50000}$ possible combinations. Consequently, Alice cannot identify the specific matrix cell currently under computation. This means that she is completely blind to the information on which matrix cell corresponds to a given sequence of editing costs without adding any computational complexity. As a result, Alice cannot identify the letter in Bob's scanpath based on the received values for the minimum cost computation. 

Additionally, we implement masking and permutation to further enhance security. In the masking process, we first decide at random whether to use order-preserving masking, and we make this decision with a probability of 0.5. Each option involves a different set of equations and this random selection strategy results in \(2^{m \times n}\) different combinations when considering the entire matrix processing phase. In the order-preserving masking phase given in Equation \ref{eq:order_preserving_masking}, we enhance the system's security by introducing a random multiplication factor. Additionally, if Alice knows only the \( \mathbf{x'} \) values but not \(\mathbf{x}\) and \( \rho_1 \), she has a system of \(n\) linear equations with \(n + 1\) unknowns (the \(n\) values of \(\mathbf{x}\) and the value of \( \rho_1 \)). Such a system is underdetermined, implying that there is no unique solution. Multiple combinations of \( \mathbf{x} \) values and \( \rho_1 \) could yield the same \( \mathbf{x'} \) values; thus, Alice is unable to definitively infer the original values.

In the initial stages of the Needleman-Wunsch algorithm, the pool of candidates might be limited, increasing the likelihood that Alice could accurately deduce the original \(x\) values. To address this vulnerability, which is most pronounced in the early phases of the algorithm, we introduce an affine transformation in \ref{step: 3}. This strategy integrates three additional random variables, further hindering Alice's ability to reverse-engineer the data successfully. Consequently, Alice is presented with a system defined by three equations but with six unknowns. This underdetermined scenario significantly amplifies the complexity of her task in determining the original \(\mathbf{x}\) values, thereby markedly boosting the masking robustness of the transformed data. In addition to the strategies mentioned, we employ a permutation, which introduces a potential of \(3!\) possible combinations at each step. Considering the entire Needleman-Wunsch algorithm, this raises the computational complexity, introducing a challenge magnified by a factor of \(6^{m \times n}\).

In summary, our framework ensures the security of Alice's and Bob's scanpath data by employing the Paillier encryption combined with a comprehensive multi-layered approach. At the end of the Needleman-Wunsch algorithm, Alice and Bob are only informed of their respective scanpath lengths and the resulting similarity score as intended.

\section{Implementation and Evaluation}
We implemented the scanpath comparison algorithm with Paillier on C++ due to its computational efficiency~\cite{fourment2008comparison,pereira2017energy}, using GMP library~\cite{Granlund12} according to ISO/IEC 18033-6~\cite{iso18033_6_2019}, and we employed the \(g=n+1\) selection given in~\cite{damgaard2001generalisation,damgaard2010generalization}. We utilized the cryptographically secure random number generator ``/dev/urandom'' which sources its randomness from hardware inputs and system events, ensuring a high degree of unpredictability by extracting entropy from these system activities. We utilized the Fisher-Yates shuffle algorithm~\cite{fisher1953statistical} for cryptographically secure permutations. Simulations were conducted on a Linux machine with an AMD EPYC 7763 64-core Processor. Alice and Bob communicated via the local host on this machine. We provide our source code publicly available.\footnote{\url{https://github.com/suleymanozdel/PrivacyPreservingScanpathComparison.git}}

In our experimental setup, we represented two distinct entities, namely Alice and Bob. Each of these parties has its own private scanpath records. For each experiment, Alice and Bob compared pairs of scanpaths from their respective datasets. To evaluate our method, we conducted tests using a synthetically generated dataset and three publicly available eye-tracking datasets, including 360em~\cite{agtzidis2019ground}, Salient360~\cite{rai2017saliency, rai2017dataset}, and EHTask~\cite{hu2021ehtask}. In the subsequent section, we provide a concise overview of each dataset and eye-tracking data encodings.

\subsection{Data Encodings and Datasets}
String representations of the eye-tracking data are primarily generated using fixations, which indicate where the gaze remains fixed over a certain amount of time. Eye-tracking datasets often provide either fixation data or raw gaze information. When such precise fixation points are not included, we pre-processed the raw data to create a string scanpath sequence following a methodology employed by~\cite{geisler2020minhash}. In the following, we outline the procedure for processing and encoding the scanpaths for further analysis.

The gaze data is quantized using a $7 \times 7$ grid over the presented stimulus if raw data is provided. Then, corresponding symbols (i.e., letters) are assigned. Our alphabet consists of both lowercase and uppercase letters. There are $52$ letters, and $49$ of them are used. Any repeated letter sequence lasting less than $100$ms is eliminated, as it is too short to be considered a fixation. The number of samples denoted as $N$, equivalent to $100$ms, varies across datasets due to different sampling rates. Any sequence with $N$ symbols or more is downsized by a factor of $N$ but is limited to only three consecutive characters at most. If the dataset directly provides fixation points, there is no need for a symbol reduction process. Thus, we applied a $7 \times 7$ grid and assigned unique letters to each grid cell. The scanpath is then encoded into strings using this mapping. Subsequently, the participants are equally distributed between Alice and Bob. Details regarding the scanpaths are given in Appendix \ref{Apppendix: Dataset Details}.

\paragraph{Datasets}\label{par: Syntehetic Dataset}
In the Salient360 dataset~\cite{rai2017saliency,rai2017dataset}, a total of $65$ stimuli were observed by $48$ participants. Each 360-degree stimulus was presented for a duration of $25$ seconds on a head-mounted display (HMD). The dataset consists of fixation points, represented by x and y coordinates on the equirectangular image. Therefore, we mapped the fixations onto a \(7 \times 7\) grid without needing a pre-processing step. The 360EM dataset, introduced by ~\cite{agtzidis2019ground}, consists of data from $13$ participants. In this dataset, participants watched $15$ 360-degree video clips with a resolution of $3840\times1920$. Each clip was approximately $1$ minute in length. Eye-tracking data were collected at a $120$ Hz sampling rate using a HMD. The dataset provided raw gaze data with x and y coordinates; therefore, we pre-processed the data to obtain a string representation of the scanpaths. In the EHTask dataset~\cite{hu2021ehtask}, data were collected from $30$ participants while viewing $15$ VR videos encompassing 360-degree perspectives. These videos were utilized for free viewing, visual search, saliency estimation, and object tracking tasks. Each video lasts $150$ seconds at a rate of $30$ frames per second, leading to a count of $4500$ frames for every video. The dataset captures gaze positions in terms of degrees: longitude values extend from $-180$ to $+180$ degrees, while latitude values vary from $-90$ to $+90$ degrees. We pre-processed the raw data to obtain scanpaths using a $7x7$ grid in degrees. In addition to using publicly available datasets, we created a random sequence of letters across various lengths, as detailed in Table \ref{tab:iteration_time}. Both Alice and Bob have ten scanpaths of identical length, and with pairwise comparison, we obtained 100 comparisons for each length. This experiment was intentionally designed to execute the protocol only for $m=n$ cases, offering a clear and comprehensive overview of our protocol's performance beyond publicly available datasets.

\subsection{Results}
We evaluated our methodology using three publicly available datasets to demonstrate its practicality and performance. Moreover, we conducted experiments to assess the effectiveness of our approach across randomly generated scanpaths of varying lengths (i.e., synthetic dataset.). These tests were performed using four distinct security parameters, denoted as \(\kappa\): \(512\), \(1024\), \(2048\), and \(3072\). We also provide the corresponding security strength for the Paillier cryptosystem, quantified in n-bits, which denotes the number of attempts required to successfully decrypt the encryption without authorization. Paillier cryptosystem with a security parameter of \(512\) offers a baseline level of security approximately equivalent to \(56\)-bit, according to the National Institute of Standards and Technology (NIST)~\cite{barker2018recommendation}, while parameters \(1024\), \(2048\), and \(3072\) correspond to security strengths of \(80\)-bit, \(112\)-bit, and \(128\)-bit, respectively. 

In our experiments, we executed our algorithm separately for each dataset. Table ~\ref{tab:combined_SPComp_mn} showcases the mean and standard deviation of the product of $m\times n$ for each dataset, providing a standardized measure of the dataset size and complexity. Moreover, the table enumerates the computation times for scanpath comparison under various security parameters, measured in seconds. The computation time can be represented as $O(mn\kappa^{\alpha})$, where $\alpha$ represents the computational impact of the security parameter. Additionally, Figure~\ref{fig:time_vs_letters} presents the aggregated results for all datasets, illustrating our protocol's computation time as demonstrating that our protocol's computation time scales as $O(mn)$ for a given $\kappa$.

The number of individual letter comparisons in each scanpath comparison equals \( m \times n \); thus, the computation time is proportional to this product. We achieved significantly low computation time using security parameters of \( 512 \) and \( 1024 \). For instance, when the product of \( m \) and \( n \) exceeds \( 10^5 \), roughly \( m = n = 315 \), the computation time for security parameter \(1024 \) takes only 75 minutes. When the security parameter was increased to \( 2048 \), providing \( 112 \)-bit security, the computation time increased to 7 hours. It rises to 22 hours with a \( 3072 \)-bit security parameter. 

{\footnotesize
\begin{figure}[ht]
    \centering
    \begin{minipage}{0.48\textwidth}
        \centering
        \captionof{table}{Scanpath comparison time and \(m \times n\) product for different datasets and security parameters.}
        \resizebox{\linewidth}{!}{
            \begin{tabular}{l l l r r}
                \toprule
            \textbf{Dataset} & \textbf{\(m \times n\) (mean \(\pm\) std)} & \textbf{$\kappa$} & \textbf{SP Comp. (s) (mean \(\pm\) std)} \\
            \midrule
            \multirow{4}{*}{Salient360} & \multirow{4}{*}{$569.4 \pm 508.4$} & 512 & $3.69 \pm 3.31$ \\
            & & 1024 & $24.9 \pm 22.7$ \\
            & & 2048 & $149.3 \pm 133.6$ \\
            & & 3072 & $460.1 \pm 391.4$ \\
            \hline
            \multirow{3}{*}{EHTask} & \multirow{3}{*}{$201,333.5 \pm 125,776.8$} & 512 & $1,130.1 \pm 704.9$ \\
            & & 1024 & $7,744.1 \pm 4,837.8$ \\
            & & 2048 & $50,297.0 \pm 3,1178.4$ \\
            & & 3072 & $132,552.7 \pm 85,561.4$ \\
            \hline
            \multirow{4}{*}{360em} & \multirow{3}{*}{$13,808.4 \pm 7,142.3$} & 512 & $75.4 \pm 38.9$ \\
            & & 1024 & $522.9 \pm 273.3$ \\
            & & 2048 & $3,404.9 \pm 1,811.2$ \\
            & & 3072 & $10,310.6 \pm 5,575.1$ \\
            \bottomrule
            \end{tabular}
            }
        \label{tab:combined_SPComp_mn}
    \end{minipage}\hfill
    \begin{minipage}{0.48\textwidth}
        \centering
        \includegraphics[width=\linewidth]{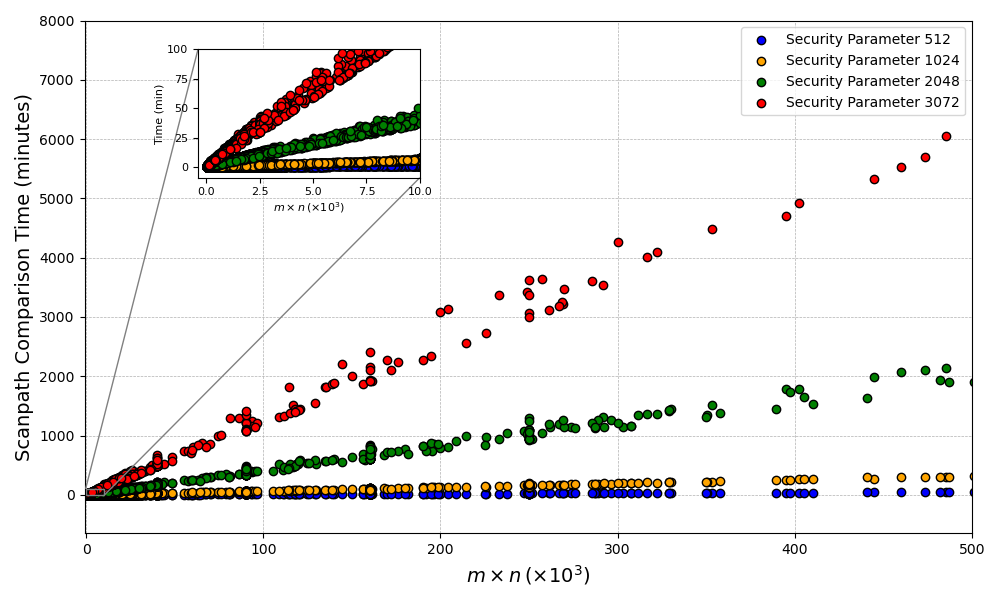}
        \caption{Scanpath comparison time vs. Letters ($m \times n$).}
        \label{fig:time_vs_letters}
    \end{minipage}
\end{figure}
}
In addition to the results from the eye-tracking datasets, which demonstrate the real-world applicability of our protocol, we have also included results from the synthetic dataset described in Section \ref{par: Syntehetic Dataset}. We carried out experiments where $m=n$, maintaining the same alphabet size of $52$ as used with the other datasets. These results are presented to provide a clearer understanding of the time requirements for the scanpath comparison task. 

{
\begin{table}[htbp]\footnotesize
\caption{Mean and standard deviation of scanpath comparison in seconds.}
\centering
\begin{tabular}{@{}|c|c|c|c|c|@{}}
\toprule 
\textbf{\(m=n\)} & \textbf{$\kappa=512$~(56-bit)} & \textbf{$\kappa=1024$~(80-bit)} & \textbf{$\kappa=2048$~(112-bit)} & \textbf{$\kappa=3072$~(128-bit)} \\
\midrule
8 & \(0.43 \pm 0.02\) & \(3.04 \pm 0.18\) & \(22.52 \pm 3.83\) & \(73.37 \pm 1.20\) \\
10 & \(0.62 \pm 0.03\) & \(4.49 \pm 0.24\) & \(32.79 \pm 5.20\) & \(108.29 \pm 1.81\) \\
20 & \(2.27 \pm 0.09\) & \(16.15 \pm 0.69\) & \(114.73 \pm 7.99\) & \(372.95 \pm 9.03\) \\
50 & \(13.92 \pm 0.67\) & \(105.08 \pm 13.18\) & \(688.61 \pm 27.69\) & \(2.27 \times 10^{3} \pm 87.83\)  \\
100 & \(58.51 \pm 7.64\) & \(401.69 \pm 30.42\) & \(2.62 \times 10^{3} \pm 136.44\) & \(8.04 \times 10^{3} \pm 860.09\) \\
200 & \(239.96 \pm 24.77\) & \(1.58 \times 10^{3} \pm 71.86\) & \(1.01 \times 10^{4} \pm 1.24 \times 10^{3}\) & \(3.18 \times 10^{4} \pm 3.29 \times 10^{3}\) \\
300 & \(542.05 \pm 36.46\) & \(3.39 \times 10^{3} \pm 243.72\) & \(2.26 \times 10^{4} \pm 2.72 \times 10^{3}\) & \(7.07 \times 10^{4} \pm 6.74 \times 10^{3}\) \\
400 & \(951.74 \pm 52.60\) & \(5.69 \times 10^{3} \pm 490.49\) & \(4.01 \times 10^{4} \pm 4.74 \times 10^{3}\) & \(1.25 \times 10^{5} \pm 1.06 \times 10^{4}\) \\
500 & \(1.49 \times 10^{3} \pm 72.35\) & \(9.47 \times 10^{3} \pm 1.03 \times 10^{3}\) & \(6.24 \times 10^{4} \pm 7.13 \times 10^{3}\) & \(1.96 \times 10^{5} \pm 1.48 \times 10^{4}\) \\
1000 & \(5.33 \times 10^{3} \pm 534.42\) & \(3.66 \times 10^{4} \pm 4.17 \times 10^{3}\) & \(2.46 \times 10^{5} \pm 1.46 \times 10^{4}\) & \(7.80 \times 10^{5} \pm 3.31 \times 10^{5}\) \\
\bottomrule
\end{tabular}
\label{tab:iteration_time}
\end{table}
}

In Figure~\ref{fig: bob_computation_time}, we further illustrate the time required for a single letter comparison, which equates to one iteration in the Needleman-Wunsch algorithm and depends solely on the security parameter. A single letter computation takes $0.037$ seconds with a 1024-bit security parameter. The time required reaches a maximum of $0.79$ seconds with a 3072-bit parameter, corresponding to 128-bit security. Bob's computation time also includes communication with Alice and the tasks of decrypting three numbers, finding the minimum, and encrypting the result. The time required for Alice's computation is also detailed in Figure~\ref{fig: alice_computation_time}. Alice's computation time approximately accounts for 25\% of the time required for one iteration. Therefore, Bob's computational load is roughly three times higher than Alice's.

\begin{figure}[htbp]
    \centering
    \begin{tabular}{@{}cc@{}} 

        \begin{minipage}{0.48\linewidth} 
            \centering
            \begin{tikzpicture}[scale=0.7]
                \begin{axis}[
                    ybar,
                    width=6cm, height=4cm, 
                    ylabel={\small{Computation Time (seconds)}},
                    ylabel near ticks, ylabel shift=-2pt,
                    symbolic x coords={512, 1024, 2048, 3072},
                    xtick=data,
                    xticklabel style={rotate=45, anchor=east},
                    bar width=14pt, 
                    enlarge x limits=0.1,
                    ymin=0, ymax=0.25,
                    ytick={0,0.05,0.1,0.15,0.20,0.25},
                    yticklabels={0,0.05,0.1,0.15,0.2,0.25}, 
                ]

                \addplot+[
                    color=black,
                    fill=gray,
                    error bars/.cd,
                    y dir=both,
                    y explicit
                ] coordinates {
                    (512, 0.001443) +- (0, 0.000164)
                    (1024, 0.009914) +- (0, 0.000851)
                    (2048, 0.063402) +- (0, 0.007192)
                    (3072, 0.206448) +- (0, 0.016885)
                };

                \end{axis}
            \end{tikzpicture}
            \caption{Min. computation time for edit distance costs for Alice across different security parameters.}\label{fig: alice_computation_time}
        \end{minipage}
        \hfill
        \begin{minipage}{0.46\linewidth} 
            \centering
            \begin{tikzpicture}[scale=0.7]
                \begin{axis}[
                    ybar,
                    width=6cm, height=4cm, 
                    ylabel={\small{Computation Time (seconds)}},
                    ylabel near ticks, ylabel shift=-2pt,
                    symbolic x coords={512, 1024, 2048, 3072},
                    xtick=data,
                    xticklabel style={rotate=45, anchor=east},
                    bar width=14pt, 
                    enlarge x limits=0.1,
                    ymin=0, ymax=1, 
                ]

                \addplot+[
    color=black, 
    fill=gray, 
    error bars/.cd, 
    y dir=both, 
    y explicit
] coordinates {
    (512, 0.005462) +- (0, 0.000625)
    (1024, 0.037187) +- (0, 0.003511)
    (2048, 0.252462) +- (0, 0.028816)
    (3072, 0.820365) +- (0, 0.059302)
};

                \end{axis}
            \end{tikzpicture}
            \caption{Cell computation time for Bob across different security parameters.}\label{fig: bob_computation_time}
        \end{minipage}
        
    \end{tabular}
\end{figure}

Further assessment of our protocol's communication overhead was performed for \( m = n = 100 \) case. The experiments resulted in total data transmissions of \( 13 \, \text{MB} \), \( 26.5 \, \text{MB} \), \( 53.2 \, \text{MB} \), and \( 79.7 \, \text{MB} \) at security levels of \( 512 \), \( 1024 \), \( 2048 \), and \( 3072 \) bits, respectively. Approximately two-thirds of the total data was sent by Bob. The findings highlight the communication efficiency of our protocol due to the Paillier cryptosystem's advantage in necessitating smaller ciphertext sizes.

\section{Discussion}
Our proposed protocol for the private comparison of scanpaths in a two-party setting enables the processing and acquisition of similarity results without disclosing any information except the lengths of the scanpaths. It enables collaboration between two different institutions, like hospitals with eye-tracking data. Distinct from prevailing two-party computation methodologies, our approach involves a one-time transmission of encrypted substitution costs for Alice's scanpath. This strategy facilitates the support of diverse substitution cost definitions and significantly reduces the communication overhead, requiring only a single round of interaction between the parties per iteration. Additionally, our probabilistic approach in processing the alignment matrix, diverging from conventional dynamic programming, enables the concealment of the current operational step, which is not possible with secret sharing-based methodologies.

We demonstrated the applicability of our protocol by evaluating it on several eye-tracking datasets, which contain eye-tracking data that can be encoded as strings, regardless of whether the data is collected from mobile or stationary systems. We further validated our protocol's utility across various datasets by analyzing scanpaths of different lengths. Results with equal-length scanpaths were provided only in the synthetic dataset to simplify understanding, though the protocol does not require identical-length inputs. Additionally, we utilized a $7\times7$ grid for experimental purposes to encode eye-tracking data as strings; however, any grid size or object-based encoding (where letters are assigned to each gaze-targeted object) can be employed. Different grid or encoding mechanism selections will primarily affect the time required to generate the substitution cost matrix on Alice's side due to the change in alphabet size. Still, the impact on the rest of the algorithm will be negligible. Additionally, the computational demand between Alice and Bob is not symmetric; Alice's computational requirements are lower than Bob's. This asymmetry allows us the flexibility to assign roles based on the computational capabilities of the parties. Furthermore, the protocol is characterized by a relatively low necessity for data transmission, obviating the need for high bandwidth capacities.

Our protocol also exhibits the capability to conform to diverse edit distance computation schemes, including the Wagner–Fischer algorithm \cite{wagner1974string}, the Smith-Waterman algorithm \cite{smith1981identification}, and the Levenshtein distance \cite{levenshtein1966binary}. This adaptability makes our protocol suitable for executing fundamental scanpath comparison works like \cite{brandt1997spontaneous} and \cite{josephson2002attention}, which utilize edit distance measures in scanpath comparison. Additionally, our protocol can privately execute ScanMatch~\cite{cristino2010scanmatch}, a well-known scanpath comparison algorithm within the eye-tracking community. The flexibility of our approach allows for the optimal selection of algorithms for specific tasks across various domains, including DNA comparison.

One inherent limitation of the Paillier cryptosystem is that ciphertext operations can sometimes exceed the defined range. Although this range is extensive, as exemplified by a key size of $\kappa = 1024$, which allows for the encryption of numbers up to $2^{1024}$, conducting addition and multiplication operations in the encrypted domain can present significant challenges. This challenge is particularly noticeable during random addition and multiplication operations. To mitigate these challenges, implementing constraints in the random generation is crucial to stay within bounds while also considering potential security vulnerabilities that may arise from this random generation. 

Furthermore, another limitation is the absence of a unique method for encoding scanpaths as string sequences. Various string representation techniques can be employed, each impacting the length of the scanpaths differently. Some methods, which result in longer scanpaths, can increase the time required for comparison. In addition to the issues mentioned earlier, various scanpath comparison techniques use representations beyond the typical string format. To accommodate all these methods while preserving privacy, there is a need for a privacy-preserving encoding approach. 

\section{Conclusion}
We proposed a secure computation protocol designed for edit distance algorithms, specifically focusing on scanpath comparison. Our two-party secure computation protocol significantly minimizes communication costs and is integrated with the Paillier encryption scheme. In future work, we aim to expand our approach to include a broader range of scanpath comparison methods. This development will involve creating privacy-preserving encoding techniques beyond the edit distance algorithm, extending their applicability to other scanpath comparison methods.

    
\begin{acks}
We acknowledge the funding by the Deutsche Forschungsgemeinschaft (DFG, German Research Foundation) - Project number 491966293. 
\end{acks}

\bibliographystyle{ACM-Reference-Format}
\bibliography{references}

\newpage
\appendix
\begin{appendices}
\newpage

\section{Sequence Diagram of a Privacy-Preserving Two-Party Computation Protocol for the Needleman-Wunsch Algorithm} \label{app:sequence_diagram}

\begin{figure}[ht]
    \centering
    \includegraphics[width=0.98\linewidth]{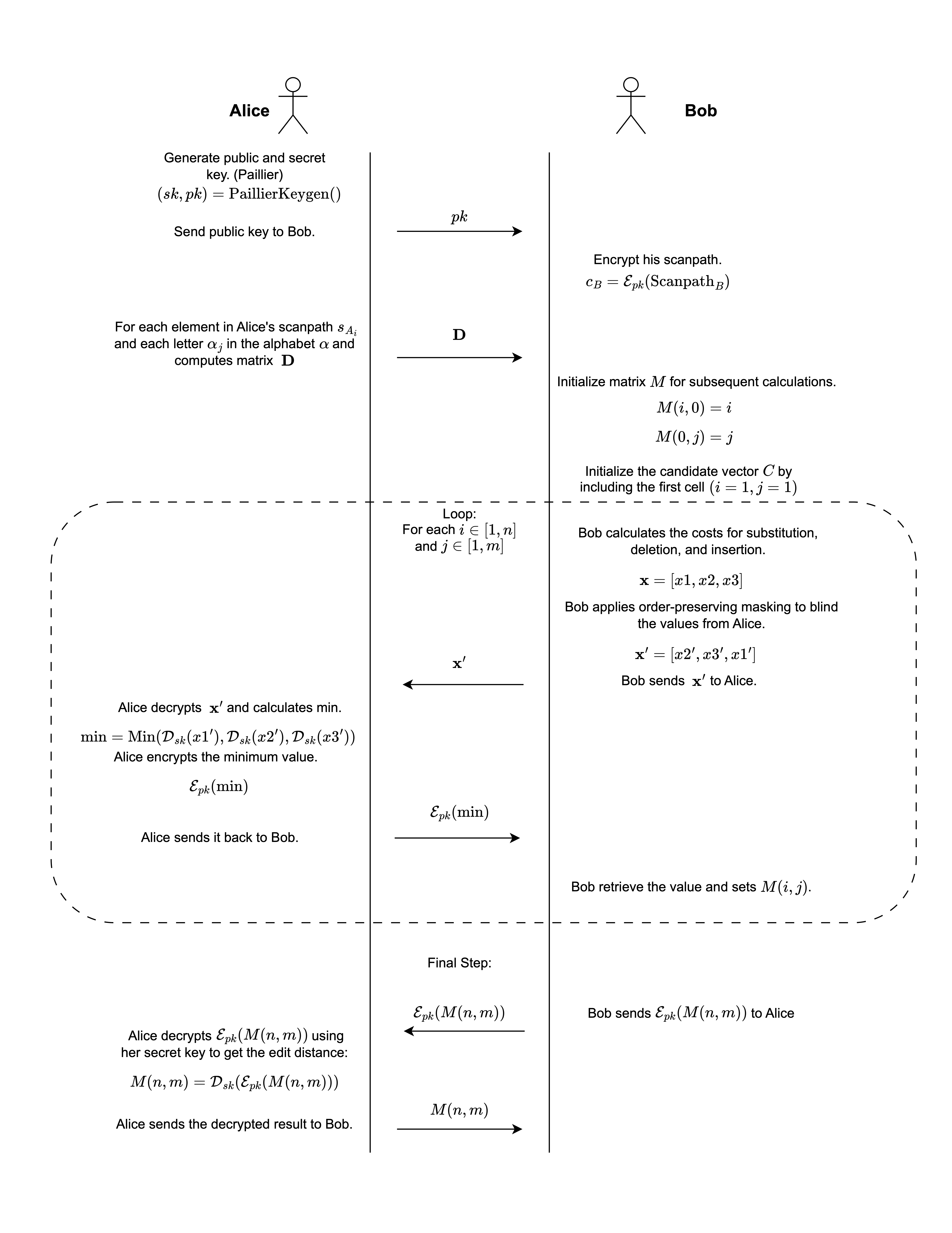}
    \caption{Sequence diagram of a privacy-preserving two-party computation protocol for the Needleman-Wunsch algorithm.}
    \Description{Sequence diagram of a privacy-preserving two-party computation protocol for the Needleman-Wunsch algorithm.}
    \label{fig:sequence_diagram}
\end{figure}

\newpage
\section{An example representation of the Matrix Processing Method}  \label{app:matrix_processing}

\begin{figure}[ht]
    \centering
    \includegraphics[width=0.9\linewidth]{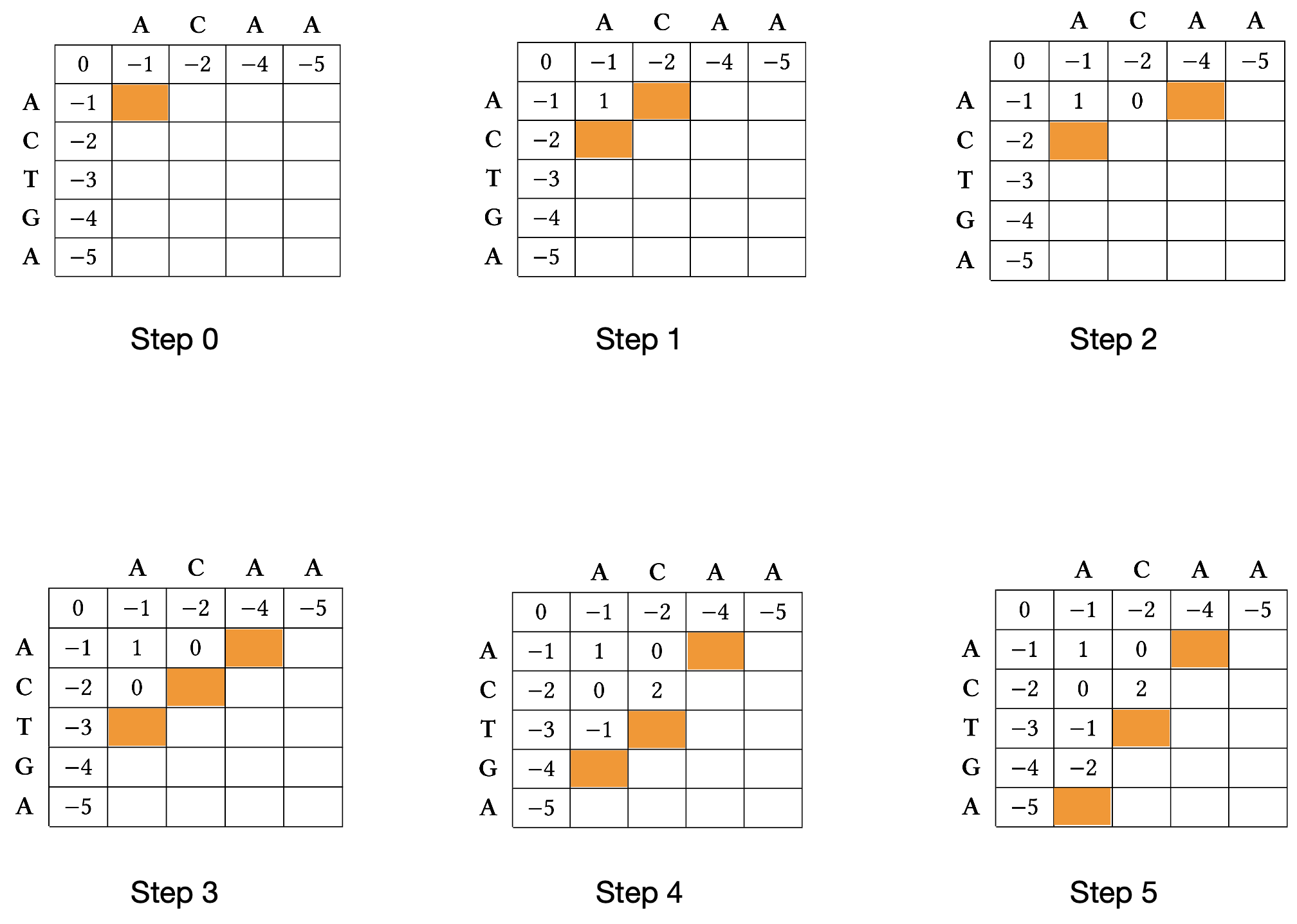}
    \caption{An example representation of the Matrix Processing Method, highlighting the cells that could be chosen in the next step (known as the candidate vector) with orange boxes.}
    \Description{An example representation of the Matrix Processing Method, highlighting the cells that could be chosen in the next step (known as the candidate vector) with orange boxes.}
    \label{fig:matrix_processing}
\end{figure}

\newpage
\section{Pseudocode of Needleman-Wunsch Algorithm} \label{app:pseudocode-NeedlemanWunch}

Below is a pseudocode representation of the Needleman-Wunsch algorithm:
\begin{algorithm}
\caption{Needleman-Wunsch Algorithm with Insertion and Deletion Costs}\label{needleman-wunsch-insertion-deletion}
\begin{algorithmic}[1]
\Procedure{NeedlemanWunsch}{$seq_1, seq_2, S, c_{ins}, c_{del}$}
    \State $m \gets$ length of $seq_1$
    \State $n \gets$ length of $seq_2$
    \State Create a 2D matrix $DP$ of size $(m+1) \times (n+1)$
    
    \For{$i \gets 0$ to $n$}
        \State $DP[i][0] \gets i \cdot c_{del}$ \Comment{Cost of deletion in $seq_1$}
    \EndFor
    
    \For{$j \gets 0$ to $m$}
        \State $DP[0][j] \gets j \cdot c_{ins}$ \Comment{Cost of insertion in $seq_2$}
    \EndFor
    
    \For{$i \gets 1$ to $m$}
        \For{$j \gets 1$ to $n$}
            \State $matchScore \gets DP[i-1][j-1] + S(seq_1[i], seq_2[j])$ \Comment{Match/Mismatch cost}
            \State $deletionScore \gets DP[i-1][j] + c_{del}$ \Comment{Cost of deletion in $seq_1$}
            \State $insertionScore \gets DP[i][j-1] + c_{ins}$ \Comment{Cost of insertion in $seq_2$}
            \State $DP[i][j] \gets \min(matchScore, insertionScore, deletionScore)$ \Comment{Fill DP matrix}
        \EndFor
    \EndFor
    
    \State \Return $DP[n][m]$ \Comment{Final alignment cost}
\EndProcedure
\end{algorithmic}
\end{algorithm}

\newpage
\section{Paillier Encryption Scheme}\label{Appendix:Paillier Encryption Scheme}
The Paillier cryptosystem is a probabilistic asymmetric additively homomorphic encryption scheme that relies on the composite residuosity class problem. This cryptosystem is particularly notable for its additive homomorphic properties. The system generates a pair of keys: a public key, typically denoted as $ pk $, and a private key, $ sk $. The private key is kept confidential by its owner and is utilized exclusively for decryption. In contrast, the public key is made available to any party wishing to encrypt data or perform other permitted operations.

\textbf{Key Generation Procedure}:
The key generation process can be summarized as 
$ (sk, pk) = \text{PaillierKeygen}(\kappa).$ where $\kappa$ which is security parameter representing the bit length of the keys.
It encompasses the following steps:

\begin{enumerate}
    \item \textbf{Key Setup:}
    \begin{itemize}
        \item Choose two distinct large prime numbers, \( p \) and \( q \). For security, these primes should be chosen randomly and remain undisclosed.
        \item Derive the modulus, \( n \), by multiplying the primes: \( n = p \times q \).
    \end{itemize}

    \item \textbf{Public Key Generation:}
    \begin{itemize}
       \item In the Paillier cryptosystem, select \( g \) as \( n + 1 \), where \( g \) is an element of the multiplicative group of integers modulo \( n^2 \), denoted by \( \mathbb{Z}_{n^2}^* \).
        \item Construct the public key as \( pk = (n, g) \).
    \end{itemize}

    \item \textbf{Private Key Generation:}
    \begin{itemize}
        \item Compute \( \lambda \) as the least common multiple of \( (p-1) \) and \( (q-1) \).
        \item Derive \( h \) by raising \( g \) to the power of \( \lambda \) modulo \( n^2 \): \( h = g^\lambda \mod n^2 \).
        \item Validate that \( h \) satisfies the condition where \( n \) divides the order of \( g \).
        \item Identify \( \mu \) as the multiplicative inverse of \( L(h) \) modulo \( n \), where \( L(x) \) is defined as \( L(x) = \frac{x-1}{n} \).
        \item Formulate the private key as \( sk = (\lambda, \mu) \).
    \end{itemize}
\end{enumerate}

Upon generating the public key \(pk\) and secret key \(sk\), the public key can be openly shared with other entities. The encryption process solely requires the public key, while the decryption process necessitates the secret key. These operations can be described as:

\begin{enumerate}
    \item \textbf{Encryption:}
    \begin{itemize}
        \item Given a message \( m \) intended for encryption, where \( 0 \leq m < n \):
         \item Randomly select an integer $ r $ from the multiplicative group of integers modulo \( n \), denoted as \( \mathbb{Z}_{n}^* \). This random selection ensures the probabilistic nature of the encryption.
        \item Encrypt \( m \) using the public key \( pk \) as:
        \[ \mathcal{E}_{pk}(m) \equiv g^m \cdot r^n \pmod{n^2} \]
    \end{itemize}

    \item \textbf{Decryption:}
    \begin{itemize}
        \item For a received ciphertext \(\mathcal{E}_{pk}(m)\), the original plaintext \( m \) is decrypted using the secret key \( sk \):
        \[ m \equiv L(\mathcal{E}_{pk}(m)^\lambda \pmod{n^2}) \cdot \mu \pmod{n} \]
    \end{itemize}
\end{enumerate}

The Paillier cryptosystem offers several notable properties, enabling arithmetic operations to be performed in the encrypted domain. These properties ensure that encrypted data remains confidential while still allowing specific computations. The properties of the Paillier cryptosystem are described below:

\begin{enumerate}

     \item \textbf{Addition of ciphertexts (\( \oplus \) operator):} Given two encrypted numbers \( \mathcal{E}_{pk}(a) \) and \( \mathcal{E}_{pk}(b) \), their encrypted sum using the \( \oplus \) operator is expressed as:
    \begin{equation}
    \mathcal{E}_{pk}(a) \oplus \mathcal{E}_{pk}(b) = \mathcal{E}_{pk}(a) \times \mathcal{E}_{pk}(b)
    \end{equation}
    And its decrypted form results in:
    \begin{equation}
    \mathcal{D}_{sk}(\mathcal{E}_{pk}(a) \oplus \mathcal{E}_{pk}(b)) = a + b
    \end{equation}
    
    \item \textbf{Scalar multiplication (\( \otimes \) operator):} For an encrypted number \( \mathcal{E}_{pk}(a) \) and a scalar \( k \), the encrypted product, when using the \( \otimes \) operator, is given by:
    \begin{equation}
    \mathcal{E}_{pk}(a) \otimes k = \mathcal{E}_{pk}(a)^k
    \end{equation}
    Decrypting this expression yields:
    \begin{equation}
    \mathcal{D}_{sk}(\mathcal{E}_{pk}(a) \otimes k) = k \cdot a
    \end{equation}
\end{enumerate}

By leveraging the homomorphic properties of the Paillier cryptosystem, along with the concept of additive and multiplicative inverses, a variety of arithmetic operations such as subtraction and division can be executed directly on ciphertexts, preserving the confidentiality of the data.
\begin{enumerate}
    \setcounter{enumi}{2} 
\item \textbf{Subtraction of Encrypted Numbers:}
Given two encrypted numbers \( \mathcal{E}_{pk}(a) \) and \( \mathcal{E}_{pk}(b) \), the subtraction operation in the encrypted domain is defined using the additive inverse of \( b \). Let \( \mathcal{E}_{pk}(-b) \) be the encryption of the additive inverse of \( b \). The encrypted difference can be computed as:
\begin{equation}
\mathcal{E}_{pk}(a) \oplus \mathcal{E}_{pk}(-b) = \mathcal{E}_{pk}(a - b)
\end{equation}
This implies that the decryption of the above result yields \( a - b \).

\item \textbf{Scalar Division on Encrypted Data:}
Given an encrypted number \( \mathcal{E}_{pk}(a) \) and a scalar \( k \), the scalar division in the encrypted domain is defined using the multiplicative inverse of \( k \) modulo \( n \). Let \( k^{-1} \) denote the multiplicative inverse of \( k \) such that \( k \times k^{-1} \equiv 1 \mod n \). The encrypted quotient is then:
\begin{equation}
\mathcal{E}_{pk}(a) \otimes k^{-1} \mod n = \mathcal{E}_{pk}\left(\frac{a}{k}\right)
\end{equation}
This implies that decrypting the result will give \( \frac{a}{k} \).

\end{enumerate}

\newpage
\section{Order Preserving Masking Proof.} \label{Appendix:Proof Order Preserving Masking}

\textbf{Proposition:} 

Let \(\mathbf{x}\) be a vector with elements \(x_i\) and \(x_j\) such that \(i \neq j\). If \(x_i < x_j\) and \(\rho_1 \geq 1\), then \(x_i' < x_j'\), where:
\begin{align}
x_i' &= \rho_1 x_i - \sum_{\substack{k=1 \\ k \neq i}}^{n} x_k \label{eq1} \\
x_j' &= \rho_1 x_j - \sum_{\substack{k=1 \\ k \neq j}}^{n} x_k \label{eq2}
\end{align}

\textbf{Proof:}  

Using the definitions from Equations \eqref{eq1} and \eqref{eq2}, we can express \(x_i'\) and \(x_j'\) as:
\begin{align}
x_i' &= \rho_1 x_i - (x_j + \sum_{\substack{k=1 \\ k \neq i, k \neq j}}^{n} x_k) \label{eq3} \\
x_j' &= \rho_1 x_j - (x_i + \sum_{\substack{k=1 \\ k \neq i, k \neq j}}^{n} x_k) \label{eq4}
\end{align}

Computing the difference between \(x_i'\) and \(x_j'\):
\begin{align}
x_i' - x_j' &= (\rho_1 + 1) (x_i - x_j) \label{eq5}
\end{align}

Given \(x_i < x_j\), we have:
\begin{align}
x_i - x_j &< 0 \label{eq6}
\end{align}

Multiplying both sides of Equation \eqref{eq6} by \( \rho_1 + 1 \) (which is positive due to \(\rho_1 \geq 1\)):
\begin{align}
(\rho_1 + 1) (x_i - x_j) &< 0 \nonumber \\
\implies x_i' - x_j' &< 0 \nonumber \\
\implies x_i' &< x_j' \label{eq7}
\end{align}

From Equation \eqref{eq7}, we conclude that if \(x_i < x_j\) and \(\rho_1 \geq 1\), then \(x_i' < x_j'\).

\newpage
\section{Pseudocode of Privacy Preserving Scanpath Comparison Protocol} \label{app:Pseudocode Privacy Preserving Protocol}

\begin{algorithm}
\caption{SecureEditDistance}
\begin{algorithmic}[1]

\Function{SecureEditDistance}{$\mathbf{s_A}$: Array, $\mathbf{s_B}$: Array, $\boldsymbol{\alpha}$: Array: Int, $c_{ins}$: Int, $c_{del}$: Int}

    \State \textit{Setup: Key Generation and Distribution}
    \State $(sk, pk) \gets \text{PaillierKeygen()}$ \Comment{Alice}
    \State SEND $pk$ TO Bob \Comment{Alice}
    \State Alice: SEND $pk$ TO Bob
    \State \textit{Initialization}
    \State $\mathbf{D} \gets \Call{InitializeMatrix}{\mathbf{s_A}, \boldsymbol{\alpha}, Encrypt, pk}$ \Comment{Alice}
    \State SEND $\mathbf{D}$ TO Bob \Comment{Alice}
    \State $\mathbf{M} \gets \Call{InitializeMatrix}{\mathbf{s_A}, \mathbf{s_B}, c_{del}, c_{ins}, pk}$ \Comment{Bob}
    \State  \textit{Edit Distance Calculation}
    \State $\text{candidates} \gets \text{set of pairs}$ \Comment{Bob}
    \State $\text{candidates.insert}(\{1, 1\})$ \Comment{Bob}
    \While{$\text{candidates} \neq \emptyset$}
        \State \textit{Local calculations:}
        \State $x_1 \gets \mathbf{M}(i-1, j-1) \oplus \mathbf{D}(i, \alpha^{-1}(\mathbf{s_B}[j-1]))$ \Comment{Bob}
        \State $x_2 \gets \mathbf{M}(i, j-1) \oplus \text{Encrypt}(pk, c_{ins})$ \Comment{Bob}
        \State $x_3 \gets \mathbf{M}(i-1, j) \oplus \text{Encrypt}(pk, c_{del})$ \Comment{Bob}

        \If{Random() < 0.5}
            \State $x_1'$, $x_2'$, $x_3'$ = \Call{ApplyOrderPreservingMasking}{$x_1$, $x_2$, $x_3$} \Comment{Bob}
        \Else
            \State $x_1' = x_1$, $x_2' = x_2$, $x_3' = x_3$ \Comment{Bob}
        \EndIf

        \State $x_1''$, $x_2''$, $x_3''$ = \Call{ApplyAffineTransformation}{$x_1'$, $x_2'$, $x_3'$} \Comment{Bob}
        \State $\mathbf{x''_\pi} =$ \Call{ApplyPermutation}{$x_1''$, $x_2''$, $x_3''$} \Comment{Bob}
        
        \State SEND $\mathbf{x''_\pi}$ TO Alice \Comment{Bob}
        
        \State \textit{Alice's Processing Step}
        \Indent
            \State $m^{*} \gets \text{Min}(\text{Decrypt}(sk, \mathbf{x''_\pi}))$ \Comment{Alice}
            \State SEND $m^{*}$ TO Bob \Comment{Alice}
        \EndIndent
        
        \State \textit{Bob's Reception and Adjustment}
        \Indent
            \State $\mathbf{M}_{ij} \gets \text{ApplyCorrection}(m^{*}, x_1, x_2, x_3, \rho_1, \rho_2, \delta_1, \delta_2)$ \Comment{Bob}
        \EndIndent
    \EndWhile

    \State \textit{Final Result Computation and Transmission}
    \State SEND $\text{Encrypt}(pk, \mathbf{M}(\text{len}(\mathbf{s_A}), \text{len}(\mathbf{s_B})))$ TO Alice \Comment{Bob}
    \State $\text{NW Distance} \gets \text{Decrypt}(sk, \text{Encrypt}(pk, \mathbf{M}(\text{len}(\mathbf{s_A}), \text{len}(\mathbf{s_B}))))$ \Comment{Alice}
    \State SEND $\text{NW Distance}$ TO Bob \Comment{Alice}

    \State \Return $\text{NW Distance}$
\EndFunction
\end{algorithmic}
\end{algorithm}

\newpage
\begin{algorithm}
\caption{Matrix Initialization Using Distances}
\begin{algorithmic}[1]
\Function{InitializeMatrix}{$s_A, \alpha, Encrypt, pk$}
    \State $D \gets \text{matrix of size } |s_A| \times |\alpha|$
    \For{each element $s_{Ai}$ in $s_A$}
        \For{each letter $\alpha_j$ in $\alpha$}
            \If{\text{some condition for absolute subtraction}} \Comment{You can specify a condition if needed}
                \State $d_{ij} \gets |s_{Ai} - \alpha_j|$  \Comment{Absolute subtraction}
            \Else
                \State $d_{ij} \gets S(s_{Ai}, \alpha_j)$  \Comment{Using predefined distances}
            \EndIf
            \State $D[i][j] \gets \text{Encrypt}(pk, d_{ij})$
        \EndFor
    \EndFor
    \State \Return $D$
\EndFunction
\end{algorithmic}
\end{algorithm}

\begin{algorithm}
\caption{Matrix Initialization for Sequence Alignment}
\begin{algorithmic}[1]
\Function{InitializeMatrix}{$s_A, s_B, c_{del}, c_{ins}, pk$}
    \State $M \gets \text{matrix of size } (|s_A|+1) \times (|s_B|+1)$
    \For{$i=0$ \textbf{to} $|s_A|$}
        \State $m_{i0} \gets i \times c_{del}$
        \State $M[i][0] \gets \text{Encrypt}(pk, m_{i0})$
    \EndFor
    \For{$j=0$ \textbf{to} $|s_B|$}
        \State $m_{0j} \gets j \times c_{ins}$
        \State $M[0][j] \gets \text{Encrypt}(pk, m_{0j})$
    \EndFor
    \State \Return $M$
\EndFunction
\end{algorithmic}
\end{algorithm}

\begin{algorithm}
\caption{SecureEditDistance}
\begin{algorithmic}[1]
\Function{ApplyOrderPreservingMasking}{$x_1$, $x_2$, $x_3$}
    \State $\rho_1 \gets \text{RandomValue()}$
    \State $x_1' \gets (x_1 \otimes \rho_1) \oplus (-x_2) \oplus (-x_3)$
    \State $x_2' \gets (x_2 \otimes \rho_1) \oplus (-x_1) \oplus (-x_3)$
    \State $x_3' \gets (x_3) \otimes \rho_1) \oplus (-x_1) \oplus (-x_2)$
    \State \Return $x_1'$, $x_2'$, $x_3'$
\EndFunction
\end{algorithmic}
\end{algorithm}

\newpage
\begin{algorithm}
\caption{SecureEditDistance}
\begin{algorithmic}[1]
\Function{ApplyAffineTransformation}{$x_1'$, $x_2'$, $x_3'$}
    \State $\rho_2 \gets \text{RandomValue()}$
    \State $\delta_1$, $\delta_2 \gets \text{RandomValues()}$
    \State $x_1'' \gets (x_1' \oplus \rho_2) \oplus \text{Encrypt}(pk, \delta_1) \oplus \text{Encrypt}(pk, \delta_2)$
    \State $x_2'' \gets (x_2' \oplus \rho_2) \oplus \text{Encrypt}(pk, \delta_1) \oplus \text{Encrypt}(pk, \delta_2)$
    \State $x_3'' \gets (x_3' \oplus \rho_2) \oplus \text{Encrypt}(pk, \delta_1) \oplus \text{Encrypt}(pk, \delta_2)$
    \State \Return $x_1''$, $x_2''$, $x_3''$
\EndFunction
\end{algorithmic}
\end{algorithm}

\begin{algorithm}
\caption{SecureEditDistance}
\begin{algorithmic}[1]
\Function{ApplyPermutation}{$x_1''$, $x_2''$, $x_3''$}
    \State $\pi \gets \text{RandomPermutation()}$
    \State $x''_\pi \gets [x''_{\pi(1)}, x''_{\pi(2)}, x''_{\pi(3)}]$
    \State \Return $x''_\pi$
\EndFunction
\end{algorithmic}
\end{algorithm}

\begin{algorithm}
\caption{Bob Correction Operation}
\begin{algorithmic}[1]
\Function{BobCorrectionOperation}{$\mathcal{E}_{pk}(m^*)$, $\delta_1$, $\delta_2$, $\rho_1$, $\rho_2$, $x_1$, $x_2$, $x_3$, \textit{maskApplied}}
    \State // Apply inverse affine transform to obtain $x_{\text{min}}'$
    \State $x_{\text{min}}' \gets (\mathcal{E}_{pk}(m^*) \oplus \mathcal{E}_{pk}(-\delta_2)) \otimes \rho_2^{-1} \oplus \mathcal{E}_{pk}(-\delta_1)$
    \State 
    \If{not \textit{maskApplied}}
        \State $M(i, j) \gets x_{\text{min}}'$
    \Else
        \State $M(i, j) \gets x_{\text{min}}' \oplus (x_1 \oplus x_2 \oplus x_3 ) \otimes (\rho_1 + 1)^{-1}$
    \EndIf
    \State \Return $M(i, j)$
\EndFunction
\end{algorithmic}
\end{algorithm}

\newpage
\section{Dataset Details}\label{Apppendix: Dataset Details}

\begin{table}[ht]
\caption{Dataset details for Alice and Bob.}
\begin{tabular}{l l r r r}
\toprule
& & \textbf{Salient360} & \textbf{EHTask} & \textbf{360em} \\
\midrule
\multirow{5}{*}{\textbf{Alice}} & Avg. \# of Scanpaths & 50.66 & 3 & 13.71 \\
& Mean & 23.743 & 416.156 & 118.948 \\
& Min & 2.00 & 95.00 & 26.00 \\
& Max & 63.00 & 849.00 & 210.00 \\
& Std Dev & 13.040 & 197.538 & 39.342 \\
\midrule
\multirow{5}{*}{\textbf{Bob}} & Avg. \# of Scanpaths & 49.37 & 3 & 14.67 \\
& Mean & 23.128 & 479.978 & 116.648 \\
& Min & 2.00 & 150.00 & 18.00 \\
& Max & 66.00 & 825.00 & 192.00 \\
& Std Dev & 12.797 & 173.409 & 44.663 \\
\midrule
\multirow{2}{*}{\textbf{Summary}} & \(m \times n\) & 569.4 & 201333.459 & 13808.419 \\
& \# Total Comparisons & 70796 & 135 & 1320 \\
\bottomrule
\end{tabular}
\end{table}

\end{appendices}
\end{document}